\documentclass[preprint, 12pt,nofootinbib]{revtex4}

\usepackage{amsfonts}
\usepackage{amsmath}
\usepackage{amssymb}
\usepackage{mathptmx}
\usepackage[dvips]{color}
\usepackage{epsfig}
\usepackage{graphicx}
\usepackage{cancel}

\def\beq{\begin{equation}}
\def\eeq{\end{equation}}
\def\be{\begin{equation}}
\def\ee{\end{equation}}
\def\bea{\begin{eqnarray}}
\def\eea{\end{eqnarray}}
\def\to{\rightarrow}

\newcommand{\calA}{{\cal A}}
\newcommand{\calL}{{\cal L}}
\newcommand{\calO}{{\cal O}}

\newcommand{\dslash}{/\!\!\!\!\!\partial}
\newcommand{\pslash}{/\!\!\!\!\!p}

\newcommand{\GeV}{{\rm GeV}}

\newcommand{\TeV}{{\rm TeV}}

\newcommand{\cm}{{\rm cm}}
\newcommand{\kpc}{{\rm kpc}}

\begin{document}

\title{Comprehensive constraints on a spin-3/2 singlet particle \\
as a dark matter candidate}
\author{Ran Ding~$^{a}$}
\email{dingran@mail.nankai.edu.cn}
\author{Yi Liao~$^{a}$}
\email{liaoy@nankai.edu.cn}
\author{Ji-Yuan Liu~$^{b}$}
\email{liujy@tjut.edu.cn}
\author{Kai Wang~$^{c}$}
\email{wangkai1@zju.edu.cn}
\affiliation{
$^a$~School of Physics, Nankai University, Tianjin 300071,
China
\\
$^b$~College of Science, Tianjin University of Technology, Tianjin
300384, China
\\
$^c$ Zhejiang Institute of Modern Physics and Department of Physics,
\\
Zhejiang University, Hangzhou, Zhejiang 310027, China
}

\begin{abstract}

We consider the proposal that dark matter (DM) is composed of a
spin-3/2 particle that is a singlet of the standard model (SM). Its
leading effective interactions with ordinary matter involve a pair
of their fields and a pair of SM fermions, in the form of products
of chiral currents. We make a comprehensive analysis on possible
phenomenological effects of the interactions in various experiments
and observations. These include collider searches for monojet plus
missing transverse energy events, direct detections of DM scattering
off nuclei, possible impacts on the gamma rays and
antiproton-to-proton flux ratio in cosmic rays, and the observed
relic density. The current data already set strong constraints on
the effective interactions in a complementary manner. The constraint
from collider searches is most effective at a relatively low mass of
DM, and the antiproton-to-proton flux ratio offers the best bound
for a heavy DM, while the spin-independent direct detection is the
best in between. For DM mass of order 10~GeV to 1~TeV, the effective
interaction scale is constrained to be typically above a few tens
TeV.

\end{abstract}

\maketitle

\section{Introduction}
\label{sec:intro}

The evidence for the domination of dark matter (DM) over ordinary
matter in our universe is robust, but still restricted to its
gravitational effects after years of efforts, from Zwicky's
suggestion in 1930's to explain the rotation curves of galaxies and
galaxy clusters to recent precision measurements on cosmic microwave
background; see \cite{Strigari:2010zz,Berera:2011zz} for brief
reviews. If some part of DM is nonbaryonic as data indicated and has
the nature of particles, it should interact weakly with ordinary
matter to cause other effects that could be observable by the means
of particle detection. Indeed, there have been many observational
and experimental activities trying to reveal various aspects of DM
particles, from direct and indirect detections to collider searches.
They have provided useful constraints on the nature of DM particles,
and may hopefully discover them in the near future.

Many new physics models contain massive neutral particles whose
stability is protected by certain exact or approximate symmetries,
and thus could serve as DM particles. Most extensively studied are
perhaps supersymmetric models; also popular are models based on
extra dimensions \cite{Kolb:1983fm}-\cite{Kong:2005hn}, and
little-Higgs models \cite{Cheng:2004yc,Birkedal:2006fz}, to mention
a few among many; see \cite{Jungman:1995df}-\cite{Feng:2010gw} for
detailed reviews. Since the basic properties of DM particles are
more or less fixed in these models, for instance, their spins,
structures and orders of magnitude of interactions with ordinary
matter, it is possible to make rather detailed predictions on their
observational effects. On the other hand, the physical relevance of
the models themselves remains to be experimentally verified.
Considering our still limited knowledge on DM particles, it is
necessary to avoid theoretical biases in exploring various
possibilities. In such a circumstance, the effective field theory
approach could be very useful \cite{Goodman:1984dc}. By assuming
basic properties of a DM particle such as its spin and mass, one
exhausts its effective interactions with ordinary matter that
respect known symmetries and are of leading order at low energies
while leaving interaction strengths as phenomenological parameters.
The physical effects can then be determined in terms of those
parameters and confronted with experimental measurements. If a DM
particle is fortunately discovered, the rough information gathered
for those properties and parameters could be employed as important
physical input in planning future facilities to reveal its
underlying dynamics.

The DM candidates of a spin zero, spin-1/2, and spin one particle
have been exhaustively studied in the literature in the framework of
effective field theory \cite{Kurylov:2003ra}-\cite{Frandsen:2011cg}.
Recently, two of us have considered the possibility that the DM
particle may have spin-3/2 \cite{Ding:2012sm} (see also Ref.
\cite{Yu:2011by} on the same suggestion). In that work, the quantum
numbers of the DM particle under the standard model (SM) gauge group
were not specified, and the constraints from direct and indirect
detection data were found similar to those for a spin-1/2 particle.
Since DM interacts very weakly with ordinary matter, it should more
naturally be a SM singlet. Here we consider this option and
investigate the constraints coming from collider measurements as
well as direct and indirect detections. A light, singlet DM particle
of spin-3/2 has also been studied earlier \cite{Kamenik:2011vy} in
rare decays of the $K$ and $B$ mesons where the particle appears as
missing energy in final states. While the kinematics of such a
particle is similar to that of a gravitino, which is also a DM
candidate in supergravity models (see \cite{Moroi:1995fs} for a
review), the interactions to be examined here are very different. A
charged spin-3/2 particle was also proposed earlier
\cite{Khlopov:2008ki} as a constituent of the so-called dark atoms.
More recently, a specific model of a spin-3/2 particle was suggested
and its direct detection examined \cite{Savvidy:2012qa}, in which
the particle is charged under the SM gauge group. The possible
relevance of spin-3/2 particles has also been considered in other
contexts, see for instance, Ref. \cite{Stirling:2011ya}, on collider
effects of a spin-3/2 top partner.

The paper is organized as follows. In the next section we consider
possible effective interactions of a singlet, spin-3/2 particle with
the standard model particles, and set up our conventions for
spin-3/2 particles. This is followed by sec \ref{sec:collider} on
the Large Hadron Collider (LHC) effects of DM particles that may
appear as missing energy in monojet events. In sections
\ref{sec:direct} and \ref{sec:indirect}, we consider, respectively,
the direct detection via DM scattering off nuclei and the indirect
detection through impacts on the cosmic rays. All of these
constraints are combined in sec \ref{sec:combined} together with
that from the observed relic density. We summarize briefly in the
last section.

\section{Effective interactions}
\label{sec:int}

The field corresponding to a particle of spin-3/2 and mass $M$ is
described by a vector-spinor, $\Psi_\mu$, with the constraint,
$\gamma^\mu\Psi_\mu=0$ \cite{Rarita:1941mf}. The free field
satisfies the equation of motion, $(i\dslash-M)\Psi_\mu=0$. The
wavefunction of such a particle with momentum $p$ and helicity
$\lambda$, $U_\mu(p,\lambda)$, can be constructed from that of a
Dirac particle and the polarization of a spin one particle in terms
of the Clebsch-Gordan coefficients \cite{Kusaka:1941}. For later
applications we will need the polarization sums for such a particle,
$P_{\nu\mu}(p)=\sum_\lambda U_\nu(p,\lambda)\bar U_\mu(p,\lambda)$,
and for an antiparticle, $Q_{\nu\mu}(p)=\sum_\lambda
V_\nu(p,\lambda)\bar V_\mu(p,\lambda)$, with $V_\mu(p,\lambda)$
being the antiparticle's wavefunction. They are known to be
\begin{eqnarray}
P_{\mu\nu}(p)&=&-(\pslash+M)\bigg(T_{\mu\nu}(p)-
\frac{1}{3}\gamma^\rho
T_{\rho\mu}(p)T_{\nu\sigma}(p)\gamma^\sigma\bigg),%
\label{eq_pol_sum}
\end{eqnarray}
and $Q_{\mu\nu}(p)=P_{\mu\nu}(p)|_{M\to -M}$, where
$T_{\mu\nu}(p)=g_{\mu\nu}-p_\mu p_\nu/p^2$ and $p^2=M^2$.

We consider the effective interactions of a spin-3/2 particle with
the SM fermions,
\begin{eqnarray}
L_L~(-1),~E_R~(-2);~Q_L~(1/3),~U_R~(4/3),~D_R~(-2/3);
\end{eqnarray}
where the number in parentheses denotes the hypercharge $Y$ which is
related to the electric charge $Q$ and third weak isospin $T^3$ by
the convention $Q=T^3+Y/2$. We start with the operators that involve
a pair of SM fermions and a pair of spin-3/2 fields. Lorentz
invariance allows for a list of fourteen independent structures
\cite{Ding:2012sm} as explicitly verified using the generalized
Fierz identities \cite{Liao:2012uj}. Demanding $\Psi_\mu$ to be a SM
singlet reduces the list to the following four:
\begin{eqnarray}
\calO_1^f&=&\bar\Psi_\mu\gamma^\alpha P_-\Psi^\mu\bar
f_L\gamma_\alpha f_L,
\nonumber%
\\
\calO_2^f&=&\bar\Psi_\mu\gamma^\alpha P_+\Psi^\mu\bar
f_R\gamma_\alpha f_R,
\nonumber%
\\
\calO_3^f&=&\bar\Psi_\mu\gamma^\alpha P_-\Psi^\mu\bar
f_R\gamma_\alpha f_R,
\nonumber%
\\
\calO_4^f&=&\bar\Psi_\mu\gamma^\alpha P_+\Psi^\mu\bar
f_L\gamma_\alpha f_L,
\label{eq_operators}
\end{eqnarray}
where $P_\pm=(1\pm\gamma_5)/2$ and $f_L$ ($f_R$) refers to any SM
doublet (singlet) fermion field. Note that the operators are
automatically flavor diagonal and that reshuffling a $\Psi_\mu$ with
an $f$ does not introduce independent operators according to
\cite{Liao:2012uj}. The corresponding effective interactions are
parameterized as
\begin{eqnarray}
\calL_\textrm{eff}&=&
+\Lambda^{-2}\sum_{f=L,Q}\big(c_1^f\calO_1^f+c_4^f\calO_4^f\big)
+\Lambda^{-2}\sum_{f=E,U,D}\big(c_2^f\calO_2^f+c_3^f\calO_3^f\big),
\label{eq_L}
\end{eqnarray}
where $\Lambda$ is the typical energy scale inducing the
interactions and $c^f_i$s are dimensionless real parameters
presumably of order one. Our later numerical analysis will be based
on this effective Lagrangian. To reduce the number of unknowns, we
follow the usual practice: we treat one operator at a time and
assume a universal $c^f$ for all relevant SM fermions.

We mention briefly some other operators that can be built out of the
spin-3/2 field and the SM fields using the approach in
\cite{Liao:2012uj}. Each of these operators violates either the
lepton or baryon number but not both, and is also forbidden if the
DM particle carries certain conserved parity. They could thus be
phenomenologically dangerous, and we will study them elsewhere. An
operator involving a single SM fermion requires an odd number of the
$\Psi_\mu$ field. With a single $\Psi_\mu$, such an operator has the
lowest possible dimension five, $(D^\mu \tilde
H)^\dagger\bar\Psi_\mu L_L$, plus one constructed with the help of
charge conjugation. Here $H$ is the Higgs doublet with $\tilde
H=\epsilon H^*$, and $D^\mu$ is the SM gauge covariant derivative.
There is no similar operator involving a quark field. It is not
possible either to form a dimension six operator involving a single
SM fermion and three $\Psi_\mu$s without including a genuinely
neutral field of neutrinos, $\nu_R$. If $\nu_R$ is indeed
introduced, there are then seven such operators according to the
results in \cite{Liao:2012uj}. An operator involving three SM
fermions contains at least a single $\Psi_\mu$, corresponding to a
dimension six operator. The situation is a bit complicated since one
can have pure-lepton, pure-quark, and mixed lepton-quark operators,
involving additional color contraction for the latter two. For
simplicity, we show here only the pure-lepton operators. Lorentz
invariance allows a complete and independent list of four
chirality-diagonal (like $\calO_{1,2}^f$) structures and four
chirality-flipped (like $\calO_{3,4}^f$) ones. But gauge symmetry of
SM singles out only the following operators,
$\epsilon^{ab}\overline{L^a_{iL}}\sigma^{\mu\nu}E_{kR}
\overline{L^b_{jL}}\gamma_\nu\Psi_\mu,~
\epsilon^{ab}\overline{(L^a_{iL})^C}\sigma^{\mu\nu}L^b_{jL}
\overline{E_{kR}}\gamma_\nu\Psi_\mu$, where $i,~j,~k$ stand for
family and $a,~b$ for the third weak isospin.

\section{Direct DM Production at LHC and its Constraints}
\label{sec:collider}

In this section, we study the collider phenomenology of spin-3/2 DM
using the effective interactions shown in sec \ref{sec:int}. Since
the DM particle is electrically and chromatically neutral, it is
completely invisible for detectors and only appears as missing
transverse energy ($\cancel{E}_{T}$) at LHC. The direct production
of DM pairs would then be completely invisible with nothing to
trigger on. For the trigger purpose, we could focus on the
production of a DM pair in association with an initial state
radiation jet or photon. In the case of a monojet plus
$\cancel{E}_{T}$ final state, the level one trigger requires that
the sum of the jet transverse momentum ($p^{j}_{T}$) and missing
transverse energy, $p^{j}_{T}+\cancel{E}_{T}$, be greater than
250~GeV or so at the LHC detectors. Studies on monojet signatures
with spin-1/2 DM effective operators have been performed extensively
\cite{Beltran:2010ww,Goodman:2010yf,Goodman:2010ku,Rajaraman:2011wf,Fox:2011pm,
Shoemaker:2011vi,Fortin:2011hv,Cheung:2012gi}. Here, we use the
latest data on search of monojet final states to constrain the
effective interactions involving spin-3/2 DM.
\begin{figure}[h]
\centering
\includegraphics[bb=23 557 550 804, width=0.70\textwidth]{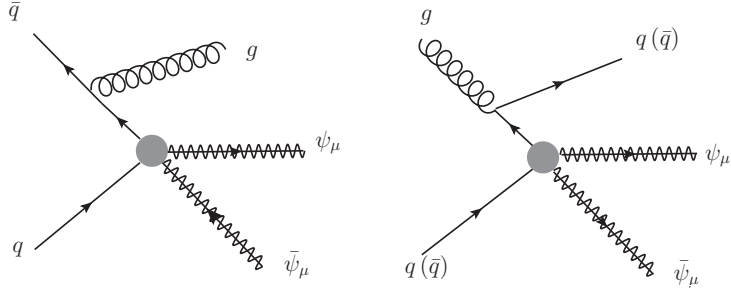}
\caption{Feynman diagrams for a monojet plus a spin-3/2 DM pair
production at a hadron collider.}
\label{fig:diagram}
\end{figure}

There are three independent subprocesses for production of a monojet
plus a DM pair,
\beq%
 q\overline{q}\to g\Psi_{\mu} \bar\Psi_{\nu} \quad \text{and}\quad
 gq (\overline{q}) \to q
 (\overline{q})\Psi_{\mu}\bar\Psi_{\nu}~,
\eeq%
corresponding to Feynman diagrams in Fig.~\ref{fig:diagram}, where
$g$ stands for a gluon. We compute in the appendix the spin- and
color-summed and -averaged matrix elements squared due to various
operators $\calO_{i}$. To illustrate the feature, we show here the
result for the subprocess $q(p_1)\bar q(p_2)\to
g(k_j)\Psi_{\mu}(k_1)\overline\Psi_{\nu}(k_2)$ due to the operator
$\calO_1$, i.e., assuming $c_1=1$ and $c_{2,3,4}=0$,
\beq%
  \sum\overline{|{\calA}|^2}=\frac{g_s^2}{9\Lambda^4}\frac{8}{9M^4 x_1 x_2}B.
  \label{eq_matrix}
\eeq%
Here $g_{s}$ is the QCD gauge coupling, and assuming massless
quarks, one has,
\bea%
 B&=&
  \Big[4 M^6 (1-x_1)^2+2 M^4 s(1-x_1-x_2)(1-x_1-y_1) (5-5 x_1-7 y_1)
\nonumber\\
   &&-4 M^2 s^2 (1-x_1-x_2)^2(1-x_1-y_1)^2
   +s^3(1-x_1-x_2)^3(1-x_1-y_1)^2\Big]
\nonumber\\
   &&+(x_1\leftrightarrow x_2,~y_1\leftrightarrow y_2),
\label{eq_subpro1}
\eea%
where $s=(p_1+p_2)^2$, and $x_{i}$, $y_{i}$ are kinematical
variables defined in the appendix. The result due to the operator
$\calO_2$ alone is identical. The same degeneracy also occurs
between the operators $\calO_3$ and $\calO_4$, and is due to the
spin summation and averaging. Our numerical analysis shows that the
difference in angular distributions between the operators
$\calO_{1,2}$ on one side and $\calO_{3,4}$ on the other, as the
expressions for various $B$s indicate, is further smeared out by
phase space integration, resulting in the same total cross sections
within about one percent. The four operators are thus equivalent for
the phenomenology considered here when a universal coupling $c_i^f$
is assumed, and from now on we will denote them simply by $\calO$.
Compared to the case of spin-1/2 DM, the production of spin-3/2 DM
is enhanced in the low mass or high energy region, and we thus
expect more stringent bounds could be set in this case.

\begin{figure}[h]
\centering
\includegraphics[width=0.70\textwidth]{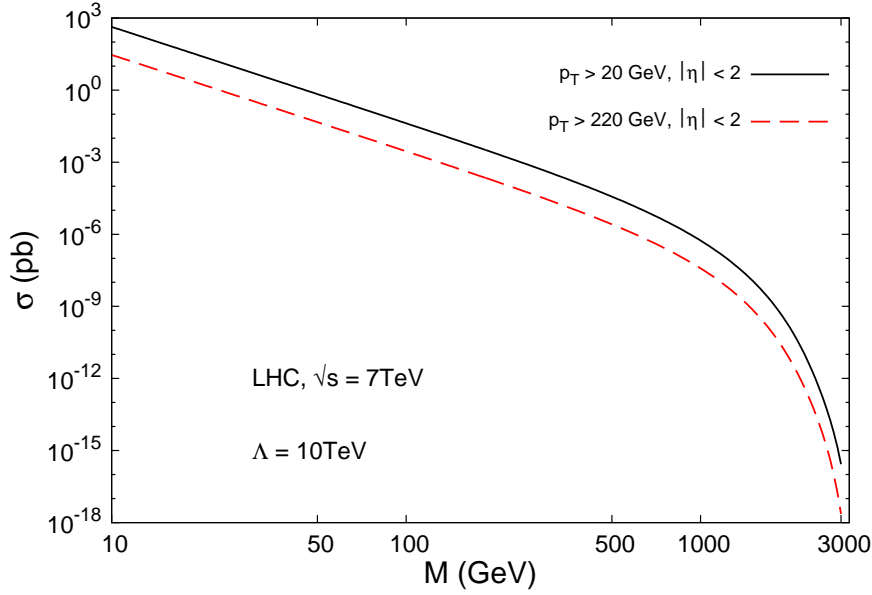}

\caption{Total partonic cross section of direct production of a DM
pair plus a monojet at 7~TeV LHC as a function of DM mass for any
one operator $\calO$. We choose the PDF set CTEQ6L1 with
renormalization scale $\mu_{R}=Q$ and factorization scale
$\mu_{F}=Q/2$, and $\Lambda=10~\TeV$.}
\label{fig:crosssection1}
\end{figure}

\begin{figure}[h]
\centering
\includegraphics[width=0.70\textwidth]{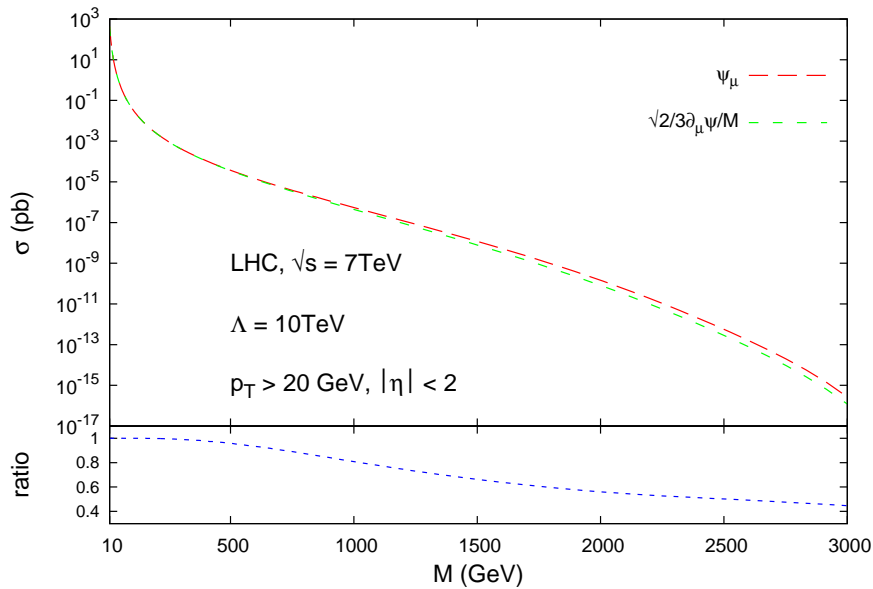}

\caption{Total partonic cross section for production of a DM pair
plus monojet at 7 TeV LHC is shown as a function of DM mass for
operator $\calO$ using the full field $\Psi_\mu$ or its effective
description, $\sqrt{2/3}M^{-1}\partial_\mu\Psi$. Also shown is the
ratio of results in the two approaches. Same physical input as in
Fig. \ref{fig:crosssection1}.}
\label{fig:crosssection}
\end{figure}

In Fig. \ref{fig:crosssection1} we show the numerical result for the
total cross section of a monojet plus $\cancel{E}_{T}$ due to any
one operator $\calO$ at 7~TeV LHC using different jet $p_{T}$ cuts.
The effective scale $\Lambda$ is fixed at 10 TeV, and we choose the
parton distribution function (PDF) set CTEQ6L1
\cite{Nadolsky:2008zw}, with renormalization scale $\mu_{R}=Q$ and
factorization scale $\mu_{F}=Q/2$. The monojet cut is within
$|\eta|_{j} < 2$ with $p_{T}^{j} > 20~{\rm GeV}$  and $p^{j}_{T} >
220~{\rm GeV}$ respectively. The cross section is dominated in the
high energy region by the $s^3$ terms in the amplitude squared; only
when the DM mass becomes non-negligible compared with $\sqrt{s}$,
does the phase space suppression begin to play a significant role
and reduce the cross section rapidly. This is reminiscent of the
well-known gravitino-goldstino equivalence in supergravity models
\cite{Fayet:1979qi,Clark:1997aa,Lee:1998aw}. In the high energy
limit, the polarization sum $P_{\mu\nu}(p)$ approaches $-\pslash
g_{\mu\nu}+ 2/(3M^2)\pslash p_{\mu}p_{\nu}$, where the first and
second term can be identified with the helicity states $\lambda =
\pm 3/2$ and $\lambda = \pm 1/2$ respectively \cite{Bolz:2000fu}.
The dominance of the latter amounts to an effective description by a
spin-1/2 field $\Psi$ via
$\Psi_\mu\to\sqrt{2/3}M^{-1}\partial_\mu\Psi$
~\cite{Mawatari:2011jy,Hagiwara:2010pi}. The results using the full
field $\Psi_\mu$ and its effective description are compared in Fig.
\ref{fig:crosssection}, where both the cross sections and their
ratio are shown. It is clear that the ratio approaches unity in the
low $M$ region and drops when $M$ increases beyond about 500~GeV.

Recently, both ATLAS \cite{ATLAS:2012ky} and CMS
\cite{Chatrchyan:2012me} collaborations have released their studies
on monojet plus $\cancel{E}_{T}$ events based on data at $\sqrt{s}=
7~\TeV$ and with an integrated luminosity of 4.7~${\rm fb}^{-1}$ and
5.0~${\rm fb}^{-1}$, respectively. We use these latest data to
derive bounds on our effective operators. Since the monojet $p_{T}$
cut is typically required to be harder than 100~GeV, we simulate our
signal events with a partonic monojet plus a DM pair without parton
showers. In addition, in order to simulate the detector performance
at the ATLAS and CMS detectors, we smear jets preformed according to
the energy resolution~\cite{:1999fq,Bayatian:2006zz}:
\begin{eqnarray}
&&\frac{\Delta E_{J}}{E_{J}} =
\frac{0.8}{\sqrt{E_{J}/\GeV}}\oplus0.15 \quad {\rm for\;\; ATLAS },
\nonumber\\
&&\frac{\Delta E_{J}}{E_{J}} = \frac{1.0}{\sqrt{E_{J}/\GeV}}\oplus0.05 \quad
{\rm for\;\; CMS }~.
\end{eqnarray}
The SM background is taken from the ATLAS/CMS analysis including the
corresponding uncertainties. The QCD jet production in principle can
contribute to monojet plus $\cancel{E}_{T}$ final states due to the
jet energy resolution. However, in this case, the distribution
$d\sigma/d \cancel{E}_{T}$ drops rapidly before $\cancel{E}_{T}<
100$~GeV. The leading  SM background then consists of a monojet plus
a $Z$ boson with invisible $Z$ decays or a monojet plus a $W^\pm$
boson decaying into soft leptons. Therefore, a large
$\cancel{E}_{T}$ with a high $p_{T}$ monojet typically works as a
good selection cut, and as we mentioned, the level one trigger is,
$p^{j}_{T} +\cancel{E}_{T}> 250~$GeV. The ATLAS/CMS searches have
assumed four sets of selection cuts, namely, SR1/SR2/SR3/SR4
\cite{ATLAS:2012ky,Chatrchyan:2012me}. In
Table~\ref{tab:monojetdata}, we summarize the selection cuts and
latest data from ATLAS and CMS.

\begin{table}[h]
\begin{center}
\renewcommand{\arraystretch}{1.2}
\begin{tabular}{|c|c|c|}
\hline
 & ATLAS $7\, \TeV$, $4.7\,{\rm fb}^{-1}$ &
   CMS $7\, \TeV$, $5.0\,{\rm fb}^{-1}$ \\
\hline
Signal region  & SR1\, /SR2\, /SR3\, /SR4 & SR1\, /SR2\, /SR3\, /SR4 \\
\hline
$\cancel{E}_{T}$ ($\GeV$) $>$ & 120\, /220\, /350\, /500 & 250\, /300\, /350\, /400 \\
\hline
$p_{T}^{j_{1}}$ ($\GeV$) $>$ & 120\, /220\, /350\, /500 & 110 \\
\hline
$|\eta|_{j_{1}}$ $<$ & 2 & 2.4 \\
\hline
$N_{\rm SM}$ & 124000\, /8800\, /750\, /83 & 7842\, /2757\, /1225\, /573 \\
\hline
$\sigma_{\rm SM}$ & 4000\, /400\, /60\, /14 & 367\, /167\, /101\, /65 \\
\hline
$N_{\rm obs}$ & 124703\, /8631\, /785\, /77 &  7584\, /2774\, /1142\, /522 \\
\hline
\end{tabular}
\end{center}
\caption{Crucial cuts and data in the ATLAS \cite{ATLAS:2012ky}
and CMS \cite{Chatrchyan:2012me} monojet plus $\cancel{E}_{T}$ analyses.
\label{tab:monojetdata}}
\end{table}

\begin{figure}[h]
\centering
\includegraphics[width=0.45\textwidth]{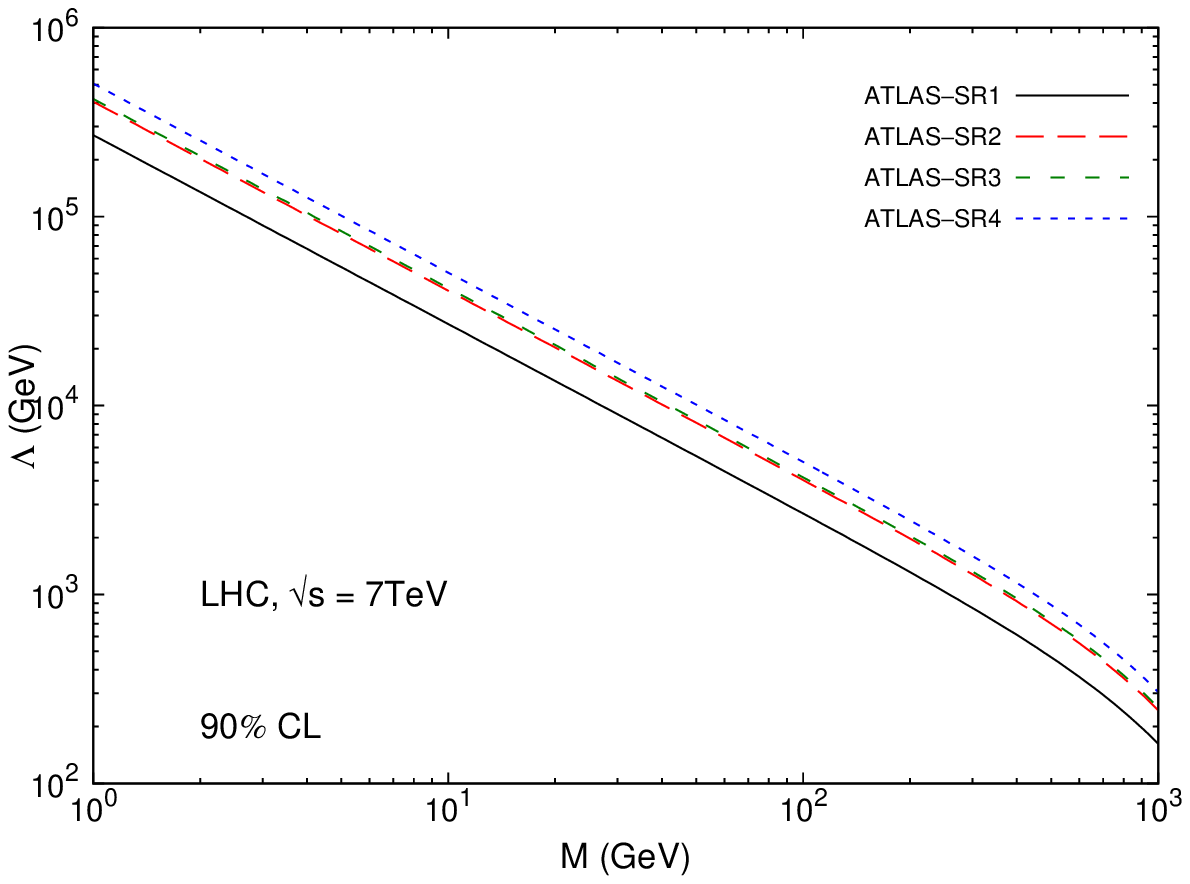}
\includegraphics[width=0.45\textwidth]{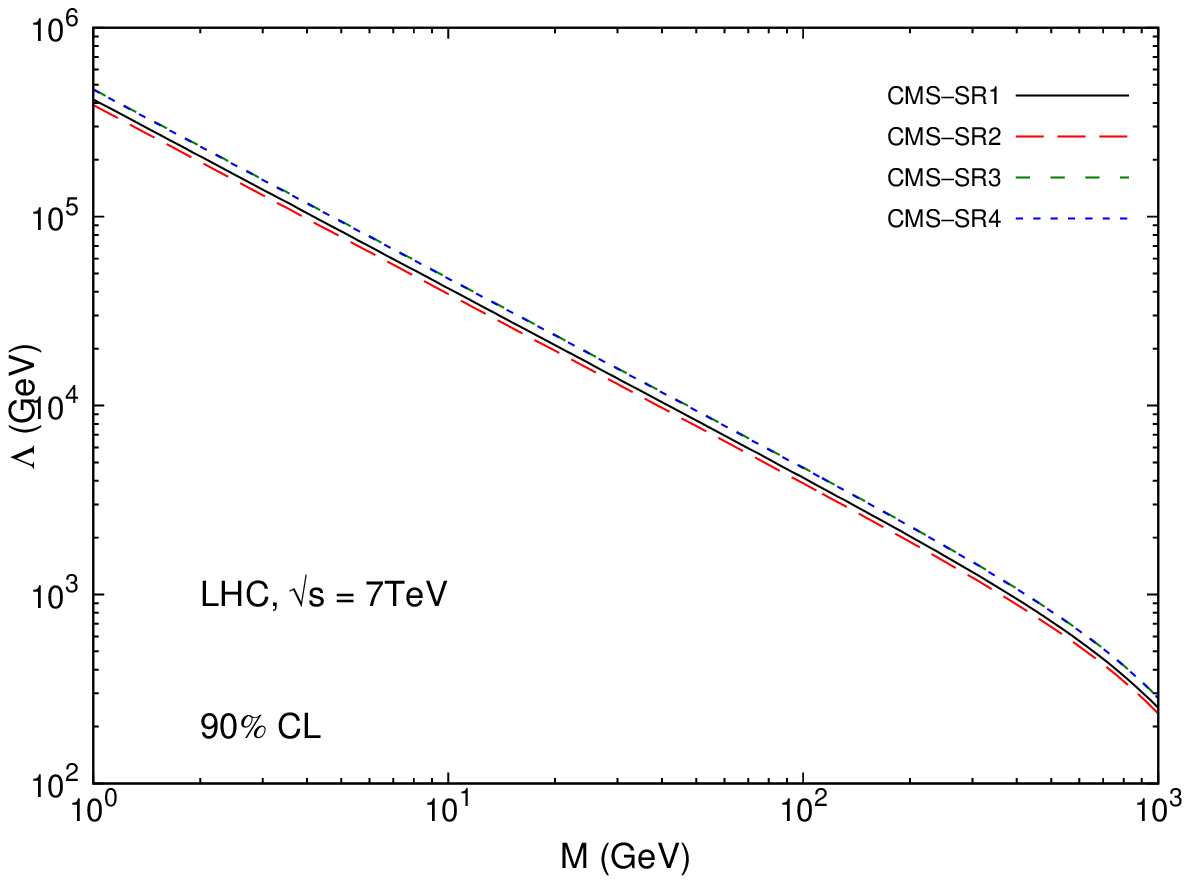}

\caption{Lower bounds on $\Lambda$ set by monojet plus
$\cancel{E}_{T}$ data from ATLAS \cite{ATLAS:2012ky} (left panel)
and CMS \cite{Chatrchyan:2012me} (right) experiments. The curves
correspond to the signal regions from SR1 to SR4 in table~\ref{tab:monojetdata}.
\label{fig:collider}}
\end{figure}

To obtain the collider bounds, we follow the $\chi^2$ definition in
Ref.~\cite{Cheung:2012gi}:
\begin{equation}
\chi^{2} = \frac{ \left( N (\Lambda,\, M) + N_{\rm SM} - N_{\rm obs}
\right )^2 } {N_{\rm obs} + \sigma_{\rm SM}^{2}} \; ,
\end{equation}
where $N_{\rm SM}$ is the number of SM background events with the
uncertainty $\sigma_{\rm SM}$ covering both statistical and
systematical uncertainties, $N_{\rm obs}$ is the number of observed
events, and $N (\Lambda,\, M)$ is the event number of DM
contribution from an effective operator. Here we require
$\chi^{2}=2.71$ to derive the 90\% CL exclusion bounds.
Fig.~\ref{fig:collider} shows the results on the effective scale
$\Lambda$ corresponding to the signal regions from SR1 to SR4,
respectively. We note that the plots have a similar shape, namely,
exhibit a nearly $M^{-4}$ slope in the light DM mass region and
begin to fall off around $M > 500~\GeV$.

\section{Direct Detection Constraints}
\label{sec:direct}

The direct detection of DM measures the event rate and energy
deposit in the collision of target nuclei ($N$) by DM particles
($\Psi_\mu$) in the local halo. It is customary to present the
results in terms of the spin-summed and -averaged cross section for
the collision at zero momentum transfer:
\begin{eqnarray}
\sigma_0=\frac{1}{16\pi(M+m_N)^2}\sum_\textrm{spins}
\overline{|\calA|^2},
\end{eqnarray}
where $\calA$ is the scattering amplitude. Nevertheless, the
procedure from the `microscopic' interactions in Eq. (\ref{eq_L}) to
the `macroscopic' scattering amplitude is nontrivial. For a nice
review on the issue with potential uncertainties incurred, see Ref.
\cite{Jungman:1995df}; for a summary of the procedure and a detailed
analysis of the operators relevant to our discussion here, see Ref.
\cite{Ding:2012sm}. Since the DM particles are nonrelativistic, they
only feel the mass and spin of a nucleus. The collision can thus be
classified into the spin-independent (SI) and spin-dependent (SD)
ones. Our operators in Eq. (\ref{eq_operators}) are linear
compositions of the 9th to 12th operators in \cite{Ding:2012sm}. In
the nonrelativistic limit, these operators are dominated by the
mass-mass and spin-spin terms, i.e., by the structures of
$\gamma^\alpha\otimes\gamma_\alpha$ and
$\gamma^\alpha\gamma_5\otimes\gamma_\alpha\gamma_5$, with the
parity-mixed terms safely ignorable. They are thus equivalent, and
each contributes simultaneously to the SI and SD cross sections:
\begin{eqnarray}
\sigma_0^\textrm{SI}&=&\frac{\mu^2}{\pi}\big(b_N\big)^2,%
\label{eq_SI}
\\
\sigma_0^\textrm{SD}&=&\frac{\mu^2}{\pi}J_N(J_N+1)\big(g_N\big)^2
\frac{20}{3},%
\label{eq_SD}
\end{eqnarray}
with $\mu=m_NM/(m_N+M)$ being the reduced mass of the $\Psi$-$N$
system. Here the effective coupling $b_N$ essentially accounts for
the contributions of valence quarks of nucleons in a nucleus of mass
number $A$ and charge $Z$:
\begin{eqnarray}
b_N=Zb_p+(A-Z)b_n,~b_p=\Lambda^{-2}(2c^u+c^d),~b_n=\Lambda^{-2}(c^u+2c^d).
\end{eqnarray}
The effective coupling $g_N$ can also be decomposed into a sum of
contributions in a nucleus of total spin $J_N$:
\begin{eqnarray}
g_N=\Lambda^{-2}\sum_qc^q\lambda_q^N,~
\lambda_q^N=J_N^{-1}\big[\langle S_p\rangle\Delta_q^p+\langle
S_n\rangle\Delta_q^n\big],
\end{eqnarray}
where $\Delta_q^{p(n)}$ is the fraction of the proton (neutron) spin
carried by the quark $q$ \cite{QCDSF:2011aa}, $\langle S_{p}\rangle$
and $\langle S_{n}\rangle$ are the expectation values of the total
spin of protons and neutrons. When we consider the DM-proton
(neutron) cross section, we have $\langle S_{p(n)}\rangle = 1/2$ and
$J_N=1/2$. For $\Delta_q^{p(n)}$, we use the values in Ref.
\cite{Belanger:2008sj}: $\Delta_{u}^{p}=\Delta_{d}^{n}=0.78\pm0.02$,
$\Delta_{d}^{p} =\Delta_{u}^{n}=-0.48\pm0.02$, and
$\Delta_{s}^{p}=\Delta_{s}^{n}=-0.15\pm0.02$. Combining
Eqs.~(\ref{eq_SI},\ref{eq_SD}), we get for any one of the operators
in Eq. (\ref{eq_operators}),
\begin{equation}
\sigma_0^{\calO}=\frac{\mu^2}{4\pi}\big[\big(b_N\big)^2
+\frac{20}{3}J_N(J_N+1)\big(g_N\big)^2\big].
\end{equation}

There exist tensions among current experimental results. Both
DAMA~\cite{Savage:2008er} and CoGENT~\cite{Aalseth:2010vx}
experiments claimed to have observed a positive signal, consistent
with a DM particle of mass $10~\GeV$ and spin-independent cross
section $\sigma_{\rm SI} \sim 2 \times 10^{-40} {\rm\, cm}^{2}$ and
$\sigma_{\rm SI} \sim 7 \times 10^{-41} {\rm\, cm}^{2}$,
respectively. On the contrary, XENON
\cite{Aprile:2012nq,Angle:2011th}, CDMS
\cite{Ahmed:2010wy,Ahmed:2009zw} and other experiments reported
negative results. We do not consider the DAMA and CoGENT results in
this paper, because one would need some isospin-violating DM model
to compromise with other experiments in a consistent way.

\begin{figure}[!htbp]
\centering
\includegraphics[width=0.70\textwidth]{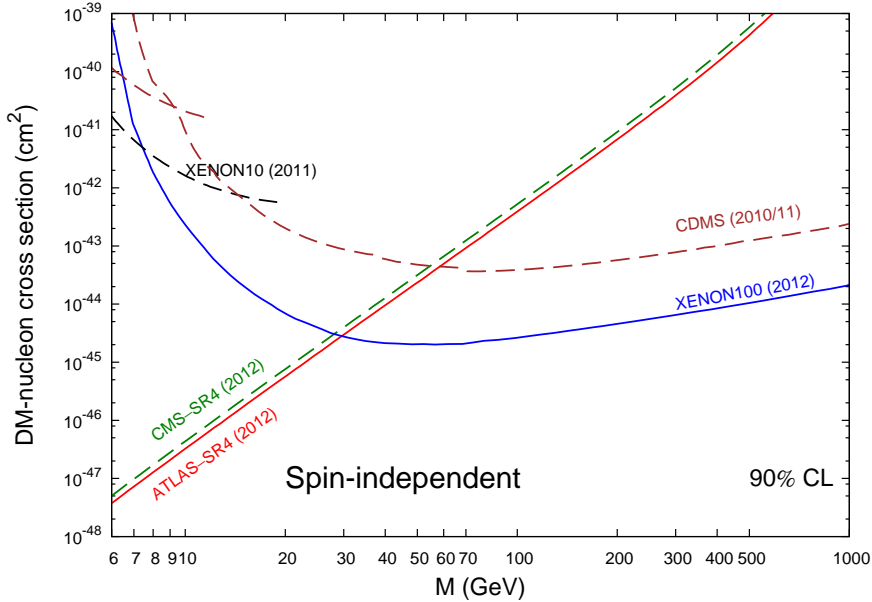}
\caption{$90\%$ CL upper limits inferred by ATLAS and CMS limits on
SI DM-nucleon cross section as a function of mass $M$. Also shown
are $90\%$ CL limits from the XENON100~\cite{Aprile:2012nq},
XENON10~\cite{Angle:2011th}, and CDMSII~\cite{Ahmed:2010wy,Ahmed:2009zw} experiments.%
\label{fig:si}}
\end{figure}

\begin{figure}[!htbp]
\centering
\includegraphics[width=0.70\textwidth]{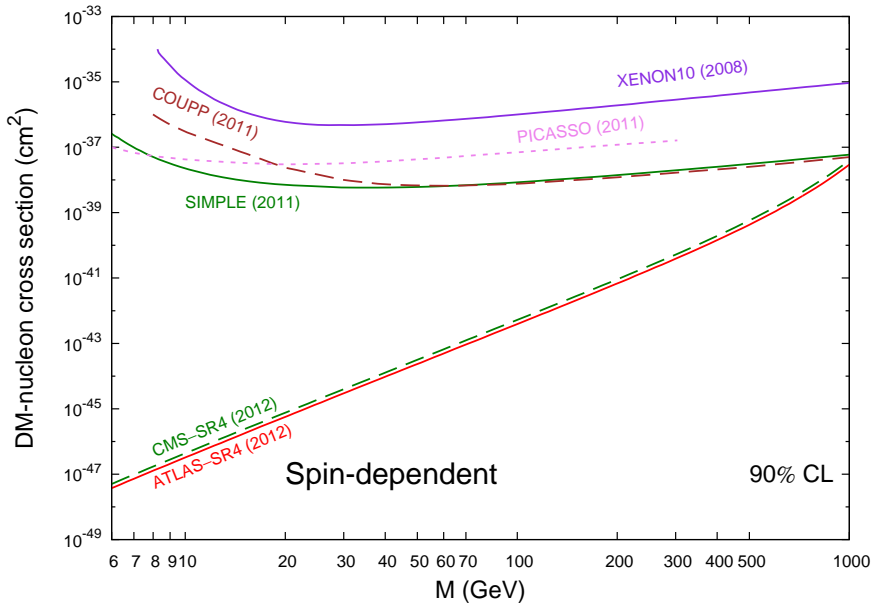}
\caption{$90\%$ upper limits inferred by ATLAS and CMS limits on SD
DM-nucleon scattering as a function of mass $M$. Also shown are
$90\%$ CL limits from the SIMPLE~\cite{Felizardo:2011uw},
COUPP~\cite{Lippincott:2011},
Picasso~\cite{Zacek:2011}, and XENON10~\cite{Angle:2008we} experiments.%
\label{fig:sd}}
\end{figure}

We present upper limits inferred by the ATLAS and CMS $90\%$ CL
limits on the SI and SD cross sections in Figs.~\ref{fig:si} and
\ref{fig:sd} respectively. For comparison, the exclusion curves from
various direct detection experiments are also depicted; see the
captions for the detail. As one can see from the figures, SD
detections generally set a weaker bound than the LHC searches, while
SI detections can yield stronger constraints than the LHC searches
for a relatively large DM mass, similarly to the case of a spin-1/2
DM particle. However, for a light DM particle of spin-3/2, the LHC
already sets a much more stringent bound than the direct detections.
This is in sharp contrast to the case of a spin-1/2 DM particle, and
originates from the enhancement discussed in sec \ref{sec:collider}.

\section{Indirect Detection Constraints}
\label{sec:indirect}

When the DM particles in our Galaxy annihilate, they produce leptons
and quarks which interact further with the interstellar medium to
initiate more secondary particles, including gamma rays, neutrinos,
positrons, antiprotons, etc. This provides an additional source of
cosmic rays on top of the known ones. Using the observed data on the
fluxes of cosmic rays, it is then possible to constrain the DM
annihilation rates. In this section, we employ the Fermi-LAT data on
the mid-latitude ($10^{\circ} < \mid b\mid < 20^{\circ}$, $0^{\circ}
< l < 360^{\circ}$) $\gamma$-rays~\cite{Abdo:2010nz} and PAMELA data
on the antiproton-to-proton flux ratio $\bar
p/p$~\cite{Adriani:2010rc} to constrain our effective interactions.

\subsection{Diffuse $\gamma$-rays and antiproton background estimation}

Gamma rays play an important role in indirect searches. As they are
not deflected by galactic magnetic fields, they point back to their
genuine sources. Several mechanisms may be responsible for the
galactic diffuse gamma ray backgrounds: decays of $\pi^{0}$ mesons
produced in nuclear interactions between cosmic rays and the nuclei
in interstellar medium, inverse Compton scattering (IC) of electrons
off photons and bremsstrahlung from electrons in the Coulomb field
of nuclei \cite{deBoer:2005tm}. In addition to the galactic
background, one also expects a contribution from the extragalactic
background (EGRB). The sources of these gamma rays include other
galaxies, unresolved point sources, large scale structures and
interactions between ultra-high energy cosmic rays and CMB photons.
Since each of them has different properties, it is difficult to
predict the shape and magnitude of EGRB. Observationally, the EGRB
can be obtained by subtraction from the data.

The positrons and antiprotons are also important indicaters of DM,
because they are relatively rare in the galactic environment. As
they traverse the interstellar space, they propagate randomly under
the influence of galactic magnetic fields, and lose energy through
processes of IC and synchrotron radiation \cite{Bertone:2004pz}.
This process may be described by a complicated diffusion equation.
The charged particles traversing the solar system are also affected
by the solar wind, which results in a shift in the spectrum observed
at the Earth compared to the interstellar one. For this, the solar
modulation potential is taken as a free parameter ranging from 300
to 1000 MV~\cite{Gleeson:1968zz,Belanger:2010gh}.

\begin{figure}[!htbp]
\centering
\includegraphics[width=0.45\textwidth]{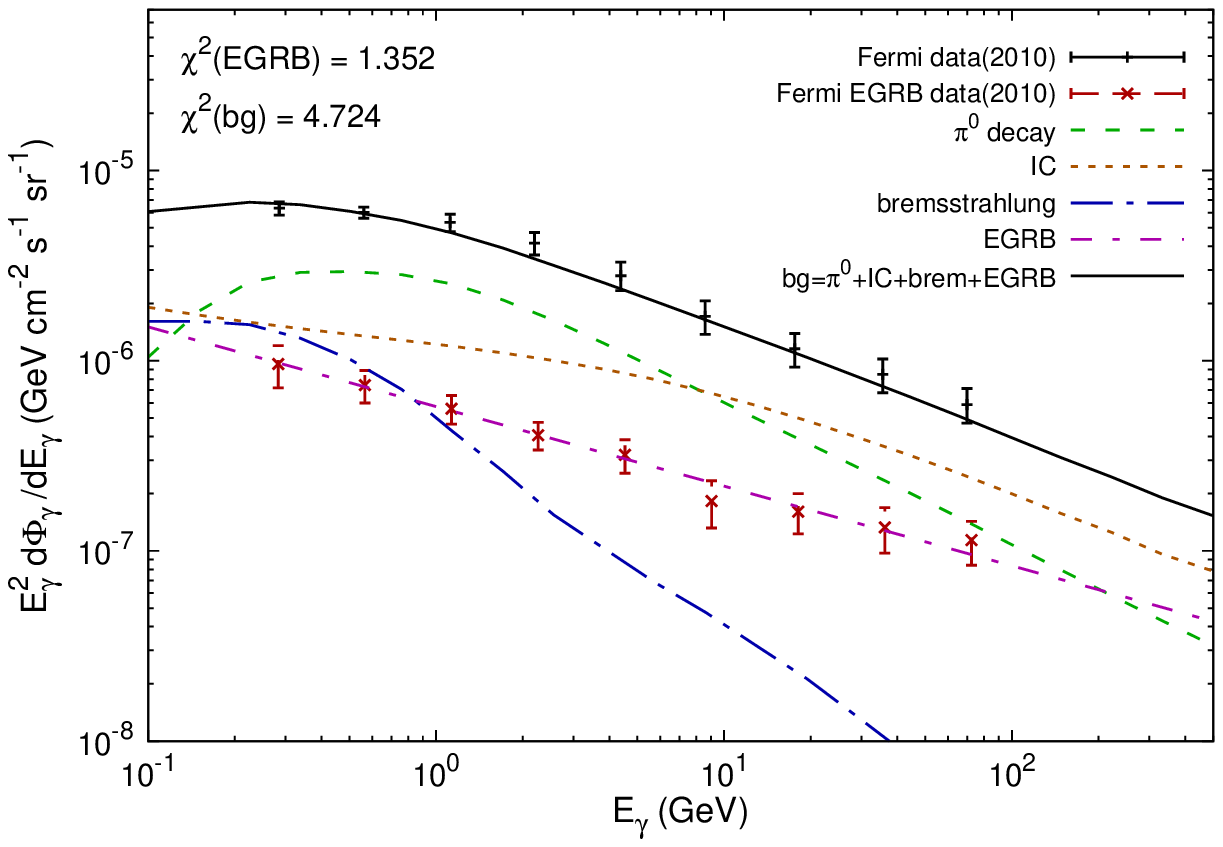}
\includegraphics[width=0.45\textwidth]{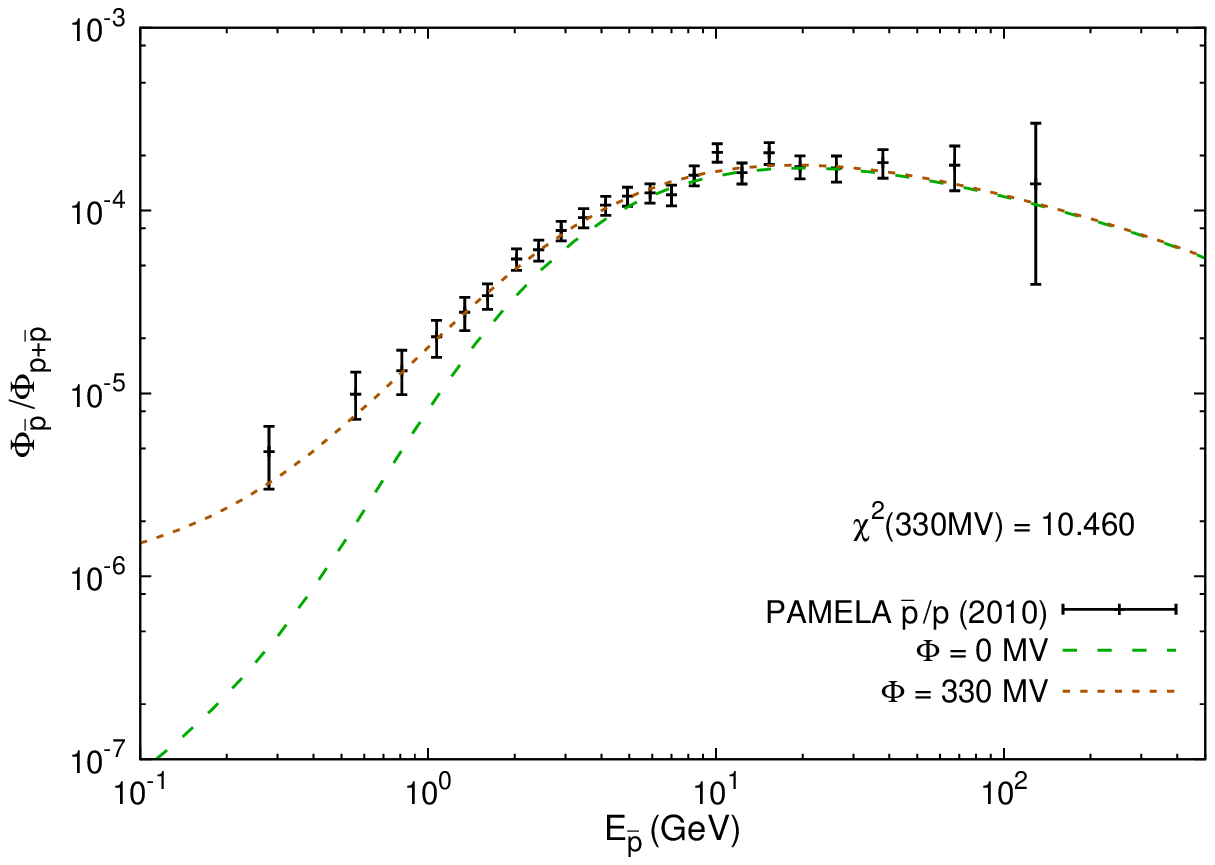}
\caption{Left panel: Background estimation for Fermi-LAT
$\gamma$-ray flux including various contributions and EGRB. The EGRB
component is obtained by fitting Fermi-LAT EGRB data with a
power-low spectrum. The background flux (dark solid line) is
obtained by adding all contributions. Right panel: Background
estimation for PAMELA $\bar p/ p$ flux ratio. The green dashed
(orange short-dashed) line corresponds to the flux without solar
modulation (with a solar modulation
potential of 330~MV).%
\label{fig:back}}
\end{figure}

A good approach to estimate the spectra of the above cosmic rays is
offered by the {\tt GALPROP} code~\cite{Strong:2009xj}. It
parameterizes the gas densities, nuclear cross sections and energy
spectra for different processes, and then solves the diffusion
equation numerically to get a complete solution for the density map
of all primary and secondary nuclei. We have therefore used {\tt
GALPROP} to calculate both background and DM annihilation components
in cosmic rays.

For the galactic diffuse $\gamma$-rays background, we calculate
various contributions and fit the Fermi-LAT EGRB data with a
power-law spectrum, $E^{2}d\Phi/dE=\Phi_{0}(E/\GeV)^{-\gamma}$. In
calculating the antiproton galactic background, we scan the solar
modulation potential to find out that the minimal $\chi^{2}$ is
reached for the PAMELA $\bar p/p$ data at a value of 330 MV. The
results are presented in Fig.~\ref{fig:back}, together with the
$\chi^{2}$ corresponding to Fermi-LAT $\gamma$-rays and PAMELA $\bar
p/p$ flux ratio data.

\subsection{Fermi-LAT and PAMELA bounds}

Now we derive indirect detection bounds using the Fermi-LAT
mid-latitude $\gamma$-rays and PAMELA $\bar p/p$ flux ratio data. As
before, we treat one operator at a time. It turns out that each of
the four operators $\calO_i$ contributes the same to the spin-summed
and -averaged total cross section for the annihilation process
$\Psi\bar\Psi\to f\bar f$,
\begin{eqnarray}
\sigma^f&=&N^f\frac{s}{16\pi\Lambda^4}\sqrt{\frac{s-4m_f^2}{s-4M^2}}~A^f.
\label{eq_anni1}
\end{eqnarray}
Here $s$ is the center-of-mass energy squared, $m_f$ and $M$ are
respectively the masses of the final ($f$) and initial ($\Psi$)
particles, and $N^f=1$ (3) when $f$ is a lepton (quark). We have set
one $c^f$ to unity and others to zero. $A^f$ is a dimensionless
function,
\begin{eqnarray}
A^f&=&\frac{1}{9} - \frac{r}{6} + \frac{1}{108\,R^2} -
\frac{r}{108\,R^2} - \frac{5}{108\,R} + \frac{2\,r}{27\,R} -
  \frac{R}{54} + \frac{8\,r\,R}{27},
\label{eq_anni2}
\end{eqnarray}
with $r=m_f^2/s$ and $R=M^2/s$. From Eqs.
(\ref{eq_anni1},\ref{eq_anni2}), one calculates the thermally
averaged annihilation cross section $\langle\sigma|v|\rangle$ to be
\begin{equation}
\langle\sigma|v|\rangle = \frac{1}{16\pi\Lambda^4} \sum_{f} N_f
\sqrt{1-\frac{m_f^2}{M^2}}\left(\frac{1}{9} \left(5 M^2+m_f^2\right)
+\frac{50 M^4-49 M^2
   m_f^2+17 m_f^4}{216 ( M^2-m_f^2 )}\langle v^2\rangle\right),
\label{eq:sv}
\end{equation}
where $v$ is the relative velocity of the annihilating DM particles,
and the average is over the DM velocity distribution in the
particular physical processes considered.

The primary particles (quarks and leptons in our case) from DM
annihilation will generate secondary particles (photons and
antiprotons) at the production point via parton showers and
hadronization. These processes can be simulated by Monte Carlo
programs. For a given DM mass from 5~TeV to 10~TeV, we use {\tt
PYTHIA6.4}~\cite{Sjostrand:2006za} to simulate
$dN_{\gamma}^{f}/dE_{\gamma}$ and $dN_{\bar p}^{f}/dE_{\bar p}$.
Here $dN_{\gamma}^{f}/dE_{\gamma}$ is the energy spectrum of photons
produced per annihilation into the final state $f$, and $dN_{\bar
p}^{f}/dE_{\bar p}$ is that of antiprotons. We considered the
$\Psi\bar\Psi\to q \bar q,~\ell\bar\ell$ channels for the photon
spectrum simulation, and $\Psi\bar\Psi\to q \bar q$ for the
antiproton. To ensure the accuracy, we took $10^6$ events in
simulation.

The photon flux in a given region $\Delta\Omega$ due to Dirac-type
DM annihilation, is written as~\cite{Cirelli:2010xx}
\begin{equation}
\frac{d\Phi_{\gamma}}{dE_{\gamma}}=\frac{1}{4\pi}\frac{\overline J
\Delta\Omega}{4 M^2}\sum_f \langle\sigma|v|\rangle_{f}
\frac{dN_{\gamma}^{f}}{dE_{\gamma}}, \quad\quad \overline J
=\int_{\Delta\Omega}d\Omega(b,l)\int_{\rm l.o.s} ds~\rho^2_{\rm halo}(r(s,\theta)),%
\label{eq gammasource}
\end{equation}
where $f$ runs over all quark and lepton channels.
$r(s,\theta)=(r_{\odot}^2+s^2-2r_{\odot}s\cos\theta)^{1/2}$ is the
galactic coordinate, $r_{\odot}$ the distance of the Sun to the
galactic center, $\theta$ the angle between directions of
observation and galactic center, and $s$ the line of sight (l.o.s)
distance. In terms of the galactic latitude $b$ and longitude $l$,
one has $\cos\theta = \cos b \cos l$. We set the integral region in
$\overline J$ to be the Fermi-LAT mid-latitude region ($10^{\circ} <
\mid b\mid < 20^{\circ}$, $0^{\circ} < l < 360^{\circ}$).  In our
calculation, we assume the NFW profile~\cite{Navarro:1995iw}:
\begin{eqnarray}
\frac{\rho_{\rm halo}(r)}{\rho_{\odot}}=\frac{r_{\odot}}{r}
\left[\frac{1+r_{\odot}/R}{1+r/R}\right]^{2},
\end{eqnarray}
where $\rho_{\odot}$ is the DM density at the solar location, and
$R$ the scale radius. We adopt the following values for these
parameters: $\rho_{\odot}=0.3~\GeV~\cm^{-3}$, $r_{\odot}=8.33~\kpc$,
and $R=20~\kpc$. The galactic DM particles should follow the
Maxwell-Boltzmann velocity distribution. We choose the velocity
dispersion $\bar v=\sqrt{\langle
v^2(r_{\odot})\rangle}=\sqrt{3/2}v_c(r_{\odot})$, with
$v_c(r_{\odot}) = 220~\rm{km}\rm{s}^{-1}$ being the local circular
velocity, and thus $\langle v^2\rangle = 2\sqrt{\langle
v^2(r_{\odot})\rangle}$. For antiprotons, the source term is
\begin{eqnarray}
Q_{\bar p}(r,E)&=&\frac{1}{4M^2}\rho^2_{\rm halo}(r)\sum_q
\langle\sigma |v|\rangle_{q}\frac{dN_{\bar p}^{q}}{dE_{\bar p}},
\label{eq pbarsource}
\end{eqnarray}
where $q$ runs over all quark channels. We then implement Eqs.
(\ref{eq gammasource},\ref{eq pbarsource}) into {\tt GALPROP} to
calculate the photon flux and $\bar p/ p$ flux ratio with the same
parameters in background estimation.

\begin{figure}[!htbp]
\centering
\includegraphics[width=0.70\textwidth]{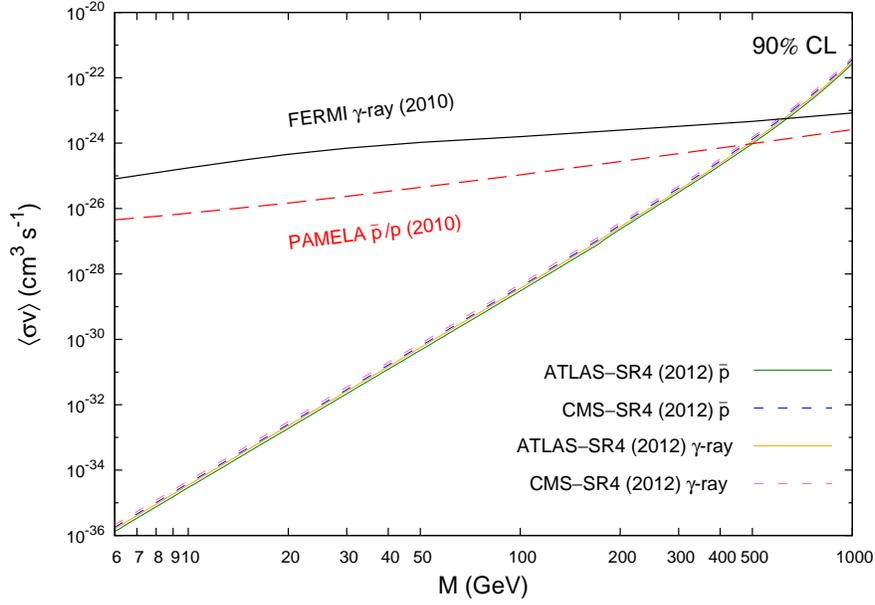}
\caption{90\% CL upper limits inferred by Fermi-LAT $\gamma$-rays
and PAMELA $\bar p/p$ flux ratio on DM annihilation cross section as
a function of mass $M$. For comparison, 90\% CL limits by ATLAS and
CMS experiments are also shown.%
\label{fig:indirect}}
\end{figure}

To obtain the exclusion bounds, we adopt a simple statistical
measurement in Ref.~\cite{Cirelli:2009dv}. Adding the DM component
with the background flux to get the total flux, $\Phi_{\rm total} =
\Phi_{\rm bkgd}+\Phi_{\rm DM}$, we define the difference
$\Delta\chi^2=\chi^2_{\rm total}(M, \langle\sigma
|v|\rangle)-\chi^2_{\rm min}$, where $\chi^2_{\rm total}(M,
\langle\sigma|v|\rangle)$ is the $\chi^2$ of the total flux and
$\chi^2_{\rm min}$ that of the background. The 90\% CL limits are
then obtained by requiring $\Delta\chi^2 = 2.71$.
Fig.~\ref{fig:indirect} shows 90\% CL limits on the thermally
averaged cross section $\langle\sigma|v|\rangle$ from the Fermi-LAT
$\gamma$-rays and PAMELA $\bar p/p$ flux ratio data. For comparison,
the upper limits from the ATLAS-SR4 and CMS-SR4 are also shown.

There exist another $\gamma$-ray limits based on the Fermi-LAT
observation of Milky Way dwarf spheroidal satellite galaxies
(dSphs)~\cite{Ackermann:2011wa}. By a joint likelihood analysis to
10 dSphs with 24 months of Fermi-LAT data, those authors obtained
the 95\% CL upper limits on (Majorana-type) DM annihilation cross
sections for several channels for a DM mass from 5~GeV to 1~TeV,
assuming DM couples only to one channel at a time. They are about
$10^{-26}-6.5\times 10^{-25}{\rm cm}^3{\rm s}^{-1}$ for the
$b\bar{b}$ channel, $8.4\times 10^{-26}-5\times 10^{-23}{\rm
cm}^3{\rm s}^{-1}$ for $\mu^+\mu^-$, and $1.4\times
10^{-26}-10^{-23}{\rm cm}^3 {\rm s}^{-1}$ for $\tau^+\tau^-$. Upon
multiplying them by a factor of two to convert to the case of
Dirac-type DM, these limits should be compared to ours on the total
cross section which has been obtained assuming that DM couples
simultaneously and equally to all possible channels. Their most
stringent limit in the $b\bar{b}$ channel is comparable to our
PAMELA $\bar p/p$ limit, while their limits in the lepton channels
$\mu^+\mu^-$ and $\tau^+\tau^-$ are comparable to our Fermi-LAT
limits but less stringent than our PAMELA $\bar p/p$ limit. One
could also use the PAMELA positron fraction
excess~\cite{Adriani:2008zr} to constrain the annihilation cross
section. However, it has been suggested that the excess might arise
from some astrophysical sources that were not accounted for earlier,
leaving the origin of excess still unclear so
far~\cite{Cirelli:2012tf,He:2009ra}.

\section{Combined Constraints}
\label{sec:combined}

In the last sections we studied the individual constraints on DM
interactions coming from LHC searches, direct and indirect
detections. In this section, we put them together with the
constraint from the DM relic density.

The currently observed relic density is a remnant of DM production
and annihilation in earlier epochs of our universe. If the
annihilation was too fast, there would be not much DM left nowadays;
and in the opposite case, DM would be over dense in the current
epoch. The observed value can therefore set a constraint on the DM
interactions responsible for annihilation. We apply the standard
procedure to calculate the relic
density~\cite{Kolbbook,Gondolo:1990dk}, i.e., by solving the
Boltzmann equation numerically with the annihilation cross section
in Eq. (\ref{eq_anni1}), summed over all fermions.

\begin{figure}[!htbp]
\centering
\includegraphics[width=0.70\textwidth]{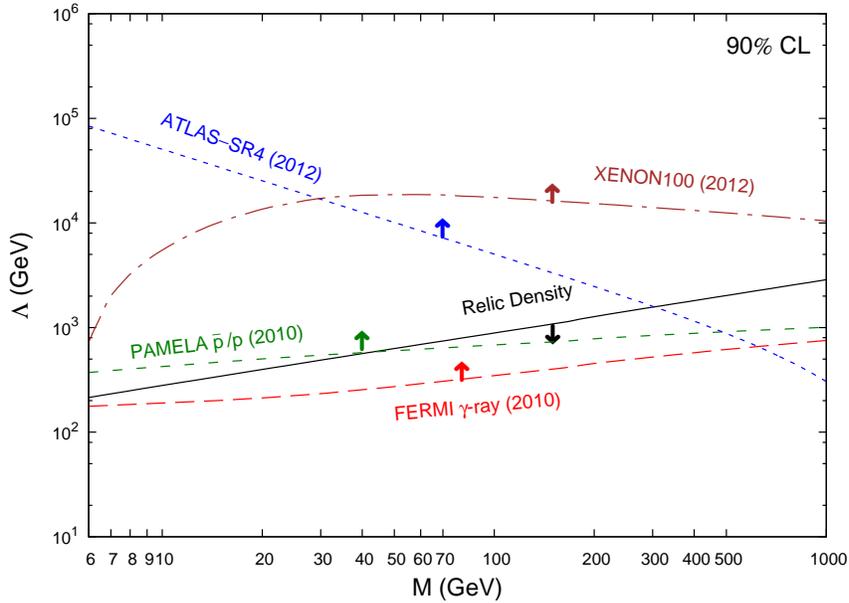}
\caption{Combined constraints on the effective scale $\Lambda$
versus DM mass from 6~GeV to 1~TeV, including observed relic
density, ATLAS experiment, SI XENON100 direct detection, Fermi-LAT
mid-latitude $\gamma$-rays data and PAMELA $\bar p/p$ flux ratio
data. The relic density bound is fixed by WMAP7+BAO+$H_0$ best-fit
value, while all other bounds correspond to 90\% lower limits.%
\label{fig:combine1}}
\end{figure}
In Fig.~\ref{fig:combine1} we present the combined constraints for
any one operator $\calO$ for DM mass from 6~GeV to 1~TeV. All limits
correspond to 90\% CL except that the relic density is fixed at
WMAP7+BAO+$H_0$ best-fit value, $\Omega_{\rm DM}h^2 =
0.1123\pm0.0035$~\cite{Komatsu:2010fb}. With one species of DM, the
latter would fix the effective scale $\Lambda$ as a function of $M$.
If instead one assumes that the particle under consideration is only
one of the DM species in the universe, the relic density will set an
upper bound on $\Lambda$. This is in contrast with all others which
set lower limits. For light DM, LHC provides the most stringent
constraint, while in the DM mass region from 30~GeV to 1~TeV, the SI
direct detection by XENON100 is most restrictive.

\begin{figure}[!htbp]
\centering
\includegraphics[width=0.70\textwidth]{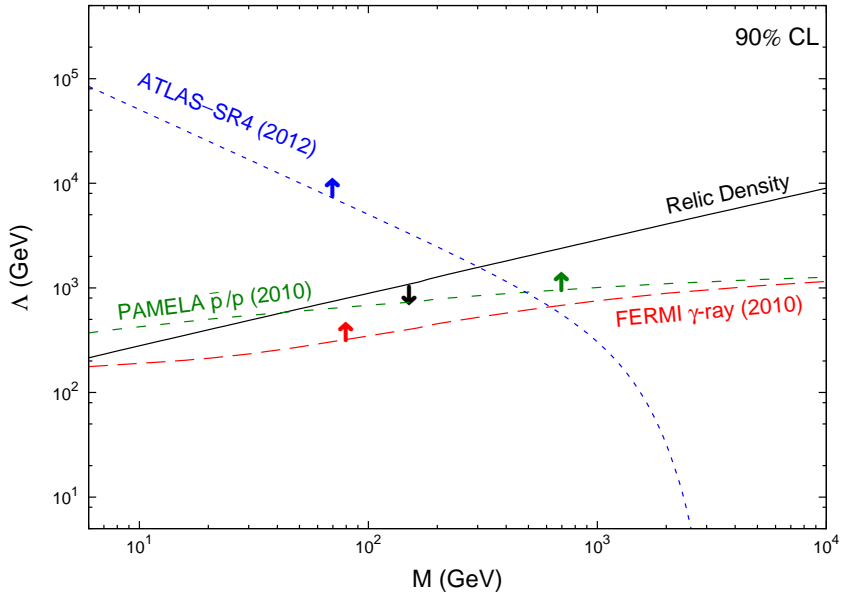}
\caption{Similar to Fig.~\ref{fig:combine1}, but for DM mass from 6~GeV to 10~TeV.%
\label{fig:combine2}}
\end{figure}

There is no direct detection constraint when the DM mass is larger
than 1~TeV. However, one can still have collider and indirect
detection constraints. In Fig.~\ref{fig:combine2}, we show the
combined constraints from the relic density, LHC searches and
indirect detections for DM mass from 6~GeV to 10~TeV. The largest DM
mass corresponding to the ATLAS limits is set to be 3~TeV. We find
that in the region $M >$ 1~TeV, the PAMELA $\bar{p}/p$ flux ratio is
most stringent. In the LHC searches, as we discussed in sec
\ref{sec:collider}, the cross section falls off rapidly in this mass
region, resulting in very weak bounds on $\Lambda$. However, as the
center of mass energy $\sqrt s$ increases, the fall-off will shift
to even larger $M$. We therefore expect that LHC running at the
projected energy of 14~TeV will be capable of pushing the bound on
$\Lambda$ further to higher values.

\section{Conclusion}

We have made a comprehensive analysis on the proposal that DM is
composed of a spin-3/2 particle which is a singlet of SM. Assuming
certain kind of parity for such a particle or assuming the lepton
and baryon numbers are still conserved, its leading effective
interactions with ordinary particles would involve a pair of them
and a pair of SM fermions. Demanding it to be a singlet turns out to
be rather restrictive. There are only four types of effective
operators in the form of products of chiral currents; and
furthermore, all of them have the same or very close
phenomenological effects thus simplifying significantly the physical
analysis.

Based on the above effective interactions we have investigated DM
effects in various experiments and observations. These include the
collider searches at LHC for monojet plus missing transverse energy
events, direct detections by spin-independent and -dependent
scattering off nuclei, indirect detections via observations on
$\gamma$-rays and antiproton-to-proton flux ratio in cosmic rays,
and the relic density. We found that the current data already set
strong and complementary constraints. For a relatively light DM
particle, say, below 30~GeV, where the DM pair production is much
enhanced, the LHC experiments provide very stringent bounds. For
instance, the latest data by ATLAS-7~TeV with an integrated
luminosity of 4.7~fb$^{-1}$ restricts the effective interaction
scale to be above 15~TeV to 100~TeV for a dark matter mass of 20~GeV
or so. The spin-independent detection by XENON100, on the other
hand, offers the most severe constraint for the mass range between
30~GeV and 1~TeV, where the effective scale is required to be above
about 20~TeV. Relatively less severely constrained is the heavy DM
scenario. For DM mass of 1-3~TeV, the strongest bound comes from the
antiproton-to-proton flux ratio in cosmic rays, which excludes
effective interactions with an effective scale lower than TeV.

\section*{Acknowledgement}

RD would like to thank the members of Zhejiang Institute of Modern
Physics for hospitality during a visit when the work was in
progress. RD and YL are supported in part by the grant NSFC-11025525
and The Fundamental Research Funds for the Central Universities
No.65030021. JYL is supported in part by the grant NSFC-11205113. KW
is supported in part by the Zhejiang University Fundamental Research
Funds for the Central Universities No.2011QNA3017 and the grants
NSFC-11245002, NSFC-11275168.

\section*{Appendix: Amplitudes squared for subprocesses of dark matter
pair production plus a monojet}

We list here the spin- and color-summed and -averaged amplitudes
squared for the subprocesses studied in sec \ref{sec:collider}. They
were used as the input file in our numerical simulation code.
Denoting the momenta of the initial-state partons, the outgoing
parton and the DM particles by $p_{1,2}$, $k_j$, and $k_{1,2}$
respectively, we define the kinematical variables
\begin{eqnarray}
 2p_a\cdot k_j=sx_a,~2p_b\cdot k_b=sy_b,
\end{eqnarray}
where $s=(p_1+p_2)^2$, $a,~b$ assume values $1,~2$, and $j$ refers
to jet. Ignoring parton masses and using $k_{1,2}^2=M^2$, we can
express other scalar products of momenta in terms of the variables,
\begin{eqnarray}
&&2p_1\cdot k_2=(1 - x_1 - y_1) s,~2p_2\cdot k_1=(1 - x_2 - y_2)s,
\nonumber
\\
&&2k_1\cdot k_j=(x_1 + y_1 - y_2)s,~2k_2\cdot k_j=(x_2-y_1+y_2)s,
\nonumber
\\
&&2k_1\cdot k_2=(1 - x_1 - x_2) s-2M^2.
\end{eqnarray}

For the sub-process $q(p_1)\bar q(p_2)\to
g(k_j)\Psi_{\mu}(k_1)\bar\Psi_{\nu}(k_2)$, the amplitude squared is
given in Eq. (\ref{eq_matrix}), where, for either of the operators
$\calO_{1,2}$, $B$ is given in Eq. (\ref{eq_subpro1}), and for
either of $\calO_{3,4}$,
\begin{eqnarray}
B&=&
  \Big[4 M^6 (1-x_1)^2+2 M^4 s(1-x_1-x_2)y_1(2 x_1+7 y_1-2)\nonumber\\
   &&\hspace{0em}-4 M^2 s^2 (1-x_1-x_2)^2y_1^2
   +s^3(1-x_1-x_2)^3y_1^2\Big]
   +(x_1\leftrightarrow x_2,~y_1\leftrightarrow y_2).
\end{eqnarray}
For the sub-process $q(p_1)g(p_2)\to
q(k_j)\Psi_{\mu}(k_1)\bar\Psi_{\nu}(k_2)$, we find
\begin{eqnarray}
\sum\overline{|\calA|^2}=\frac{g_s^2}{24\Lambda^4}\frac{8}{9M^4x_2}B,
\label{eq_subpro2}
\end{eqnarray}
where for $\calO_{1,2}$,
\begin{eqnarray}
B&=&4 M^6 [(1-x_1)^2+(x_1+x_2)^2]
  +2 M^4s (1-x_1-x_2) (24 x_1 y_1-12 x_1 y_2\nonumber\\
  &&-2 x_2 y_1+2 x_2 y_2+10 x_1^2-2
   x_2 x_1-10 x_1+14 y_1^2+7 y_2^2-12 y_1-14 y_1y_2+5)\nonumber\\
   &&-4 M^2 s^2\left(1-x_1-x_2\right){}^2
   \left(4 x_1 y_1-2 x_1 y_2+2 x_1^2-2 x_1+2 y_1^2+y_2^2-2
   y_1-2 y_1 y_2+1\right)\nonumber\\
   &&+s^3 \left(1-x_1-x_2\right){}^3
   \left(4 x_1 y_1-2 x_1 y_2+2 x_1^2-2 x_1+2 y_1^2+y_2^2-2
   y_1-2 y_1 y_2+1\right),
\end{eqnarray}
and for $\calO_{3,4}$,
\begin{eqnarray}
B&=&4 M^6 [(1-x_1)^2+(x_1+x_2)^2]
  +2 M^4s (1-x_1-x_2) (24 x_1 y_1-12 x_1 y_2\nonumber\\
  &&-2 x_2 y_1+2 x_2 y_2+10 x_1^2-2
  x_2 x_1-10 x_1+14 y_1^2+7 y_2^2-12 y_1-14 y_1y_2)\nonumber\\
  &&-4 M^2 s^2\left(1-x_1-x_2\right)^2
  \left(4 x_1 y_1-2 x_1 y_2+2 x_1^2-2 x_1+2 y_1^2+y_2^2-2
  y_1-2 y_1 y_2\right)\nonumber\\
  &&+s^3 \left(1-x_1-x_2\right)^3
  \left(-2 x_2 y_1+2 x_2 y_2+x_2^2+2 y_1^2+y_2^2-2 y_1 y_2\right).
\end{eqnarray}
Finally, the result for the subprocess $\bar q(p_2)g(p_1)\to \bar
q(k_j)\Psi_{\mu}(k_1)\bar\Psi_{\nu}(k_2)$ can be obtained from Eq.
(\ref{eq_subpro2}) by the interchanges, $x_1\leftrightarrow
x_2,~y_1\leftrightarrow y_2$.

\end{document}